\documentclass[conference]{IEEEtran}
\IEEEoverridecommandlockouts %To add sponsors and index terms

\usepackage{cite}
\ifCLASSINFOpdf
   \usepackage[pdftex]{graphicx}
   \graphicspath{{./Images/}}
   \DeclareGraphicsExtensions{.pdf,.jpeg,.png}
\else
\fi
\usepackage[cmex10]{amsmath}

\usepackage[hidelinks]{hyperref}
\usepackage{multirow}
\usepackage{booktabs,colortbl}
\usepackage{amsmath}

\begin{document}
\IEEEoverridecommandlockouts
\IEEEpubid{\makebox[\columnwidth]{{978-1-4673-8463-6/16/\$31.00~\copyright~2016 IEEE} \hfill}
\hspace{\columnsep}\makebox[\columnwidth]{ }}

%
% paper title
% can use linebreaks \\ within to get better formatting as desired
\title{Forecasting Framework for Open Access Time Series in Energy}

% author names and affiliations
% use a multiple column layout for up to two different
% affiliations
\author{%

\IEEEauthorblockN{Gergo Barta, Gabor Nagy, Gyozo Papp}\\
\IEEEauthorblockA{\emph{Department of Telecommunications and Media Informatics}\\
Budapest University of Technology and Economics\\
Budapest, Hungary\\
\{barta, nagyg, pappgy\}@tmit.bme.hu}
\and
\IEEEauthorblockN{Gabor Simon}\\
\IEEEauthorblockA{Dmlab Ltd.\\
Budapest, Hungary\\
simon.gabor@dmlab.hu}

% use \thanks{} to gain access to the first footnote area
% a separate \thanks must be used for each paragraph as LaTeX2e's \thanks
% was not built to handle multiple paragraphs
%

% \thanks{Identify applicable sponsor/s here. \emph{(sponsors)}}%

}

% use for special paper notices
%\IEEEspecialpapernotice{(Invited Paper)}

% make the title area
\maketitle

%ABSTRACT
\begin{abstract}
In this paper we propose a framework for automated forecasting of energy-related time series using open access data from European Network of Transmission System Operators for Electricity (ENTSO-E). The framework provides forecasts for various European countries using publicly available historical data only. Our solution was benchmarked using the actual load data and the country provided estimates (where available). We conclude that the proposed system can produce timely forecasts with comparable prediction accuracy in a number of cases. We also investigate the probabilistic case of forecasting - that is, providing a probability distribution rather than a simple point forecast - and incorporate it into a web based API that provides quick and easy access to reliable forecasts.
\end{abstract}

%INDEX TERMS

\section{Introduction}
There is a great deal of interest in forecasting energy-related time series, specifically regarding the generation of renewable energy sources, load forecasting, electricity price forecasting or gas consumption forecasting \cite{Aggraval, Mangalova, Gergo}. For transparency purposes two European platforms offer open access to such data: ENTSO-E and ENTSOG. The data from these sources are rich, some data quality issues arise, however, that need some investigation for the platform to be useful. In this paper we utilize the free data sources to show that releasing data to the public is beneficial and should be promoted. By building a framework for the automatic prediction of energy-related time series published on an open portal we wish to add our part of encouragement. 

Furthermore we incorporate the idea promoted in the Global Energy Forecasting Competition $2014$ \cite{Gefcom2014}; a new and unique approach to forecasting hourly renewable energy production, electric load and price was introduced: the forecasting of a probability distribution of the target variable in contrast to forecasting a single value. Probabilistic forecasts provide more detailed insight for stakeholders offering immediate business value. Our solution achieved good results in terms of pinball loss in the competition, and we assess if the framework is suitable for reliable probabilistic forecasts in a rolling fashion for a number of energy-related time series. 

The document is organized as follows. After some basic information on the used data source
in Section II, we present in Section III our comprehensive report on data quality of the platform. A brief methodological overview can be found in Section IV and the extensive tour of the internal framework structure in Section V. We perform an in-depth analysis of
real-world data, then present and discuss detailed experimental results (Section VI). Conclusions and further ideas are provided in the final Section VII.

\section{Data source}
ENTSO-E, the European Network of Transmission System Operators, represents $41$ electricity transmission system operators (TSOs) from $34$ countries across Europe. ENTSO-E was established and given legal mandates by the EU’s Third Legislative Package for the Internal Energy Market in 2009, which aims at further liberalizing the gas and electricity markets in the EU. The main objectives of ENTSO-E centre on the integration of renewable energy sources such as wind and solar power into the power system, and the completion of the internal energy market, which is central to meeting the European Union’s energy policy objectives of affordability, sustainability and security of supply\cite{entsoe}.

The ENTSO-E Transparency Platform was launched in early 2015. It is aimed at providing free, continuous access to pan-European electricity market data for all users, across categories such as load, generation, transmission, balancing, outages and congestion. Our work focuses on the country load data published on this new portal. 

\section{Data quality report}
We performed the data validation on the available load data on the ENTSO-E Transparency Platform. The time span of the validation started on January 2015 and went on until October 2015, spanning 3 seasons of the year. The portal is used to access historical load data in a timely manner. Every day we perform a screen scrape of the data source for the last day of available data. Observation of the data showed that there are $3$ major types of data reporting frequencies in the publicly available load data: hourly, half-hourly and quarter hourly data reporting.

%Data consistency will be measured by $2$ simple heuristics:
%\begin{enumerate}
  %\item did the country on the time of the download provide an "N/A"
  %\item is this value $0$
%\end{enumerate}

Data consistency was measured by $2$ simple heuristics; check if the country provided data was either 'N/A' or '0' at the time of the download. We summarize the occurrence of the above events for both of the provided data columns: Day-ahead Total Load Forecast, Actual Total Load. To get a bigger picture and a general score of data quality for each country, we observe the frequency of errors relative to the theoretical maximum reportable values for the validation time interval (Table \ref{tab:dataqual}). 

We selected a number of countries for investigation (boldfaced) based on their data quality and market size; Belgium, Czech Republic, Finland, France, Germany, Hungary, Italy, Norway, Poland and Slovakia. There are several countries that do not have any data on the platform at all in the validation time span, we omit these from further analysis. These countries are the following: Bosnia and Herzegovina, Cyprus, Belarus, Iceland, Moldavia, Malta, Russia, Turkey and Ukraine.

% Please add the following required packages to your document preamble:
% \usepackage{booktabs}
\begin{table}[]
\centering
\caption{Overall data quality assessment score}
\label{tab:dataqual}
\begin{tabular}{@{}llll@{}}
\toprule
Country code     & Country name   & Data quality & Frequency      \\ \midrule
10YAL-KESH-----5          & Albania                 & 58.5\%          & Quarter-hourly          \\
10YAT-APG------L          & Austria                 & 98.5\%          & Quarter-hourly          \\
\textbf{10YBE----------2} & \textbf{Belgium}        & \textbf{91.2\%} & \textbf{Quarter-hourly} \\
10YCA-BULGARIA-R          & Bulgaria                & 46.7\%          & Hourly                  \\
10YHR-HEP------M          & Croatia                 & 97.4\%          & Hourly                  \\
\textbf{10YCZ-CEPS-----N} & \textbf{Czech Republic} & \textbf{98.2\%} & \textbf{Hourly}         \\
10Y1001A1001A65H          & Denmark                 & 91.6\%          & Hourly                  \\
10Y1001A1001A39I          & Estonia                 & 97.2\%          & Hourly                  \\
\textbf{10YFI-1--------U} & \textbf{Finland}        & \textbf{98.8\%} & \textbf{Hourly}         \\
\textbf{10YFR-RTE------C} & \textbf{France}         & \textbf{97.7\%} & \textbf{Hourly}         \\
\textbf{10Y1001A1001A83F} & \textbf{Germany}        & \textbf{83.9\%} & \textbf{Quarter-hourly} \\
10YGR-HTSO-----Y          & Greece                  & 97.6\%          & Hourly                  \\
\textbf{10YHU-MAVIR----U} & \textbf{Hungary}        & \textbf{98.8\%} & \textbf{Quarter-hourly} \\
10YIE-1001A00010          & Ireland                 & 49.8\%          & Half-hourly             \\
\textbf{10YIT-GRTN-----B} & \textbf{Italy}          & \textbf{77.1\%} & \textbf{Hourly}         \\
10YLV-1001A00074          & Latvia                  & 98.5\%          & Hourly                  \\
10YLT-1001A0008Q          & Lithuania               & 96.8\%          & Hourly                  \\
10YLU-CEGEDEL-NQ          & Luxembourgh             & 95.7\%          & Quarter-hourly          \\
10YMK-MEPSO----8          & Macedonia               & 63.3\%          & Hourly                  \\
10YCS-CG-TSO---S          & Montenegro              & 91.3\%          & Hourly                  \\
10YNL----------L          & Netherlands             & 98.1\%          & Quarter-hourly          \\
\textbf{10YNO-0--------C} & \textbf{Norway}         & \textbf{88.7\%} & \textbf{Hourly}         \\
\textbf{10YPL-AREA-----S} & \textbf{Poland}         & \textbf{98.0\%} & \textbf{Hourly}         \\
10YPT-REN------W          & Portugal                & 98.2\%          & Hourly                  \\
10YRO-TEL------P          & Romania                 & 92.8\%          & Hourly                  \\
10YCS-SERBIATSOV          & Serbia                  & 85.8\%          & Hourly                  \\
\textbf{10YSK-SEPS-----K} & \textbf{Slovakia}       & \textbf{98.4\%} & \textbf{Hourly}         \\
10YSI-ELES-----O          & Slovenia                & 96.4\%          & Hourly                  \\
10YES-REE------0          & Spain                   & 98.6\%          & Hourly                  \\
10YSE-1--------K          & Sweden                  & 79.4\%          & Hourly                  \\
10YCH-SWISSGRIDZ          & Switzerland             & 82.4\%          & Hourly                  \\
GB                        & United Kingdom          & 48.1\%          & Half-hourly      \\ \bottomrule
\end{tabular}
\end{table}

In general we can say that in the data quality assessment period the majority 21 countries out of 32 have less than $10\%$ bad data frequency. After cross-referencing the missing values from our historical real-time database, with the state of the database of the Transparency Platform, we see that the majority of these values are backfilled and the errors are corrected. However if one tries to rely on the timely updates of the Transparency Platform one could get into trouble with missing load values and forecasts as well. Surprisingly the better reporting countries are from the Central-Eastern Europe, most notably Croatia, Poland, Slovakia and Hungary.

All of the above countries have missing values in both the forecasted load and the actual reported load, however only a handful have zeros (Croatia, Portugal, Lithuania, Montenegro, Sweden, Serbia, Macedonia and Ireland). It should be self-evident not to report zeros as forecasts or load values, we strongly advise a check to be made at the time of the report. Table \ref{tab:dataqual2} summarizes the two metrics as percentages for each country.

% Please add the following required packages to your document preamble:
% \usepackage{booktabs}
\begin{table}[]
\centering
\caption{Observed data issues in countries with sufficient data}
\label{tab:dataqual2}
\begin{tabular}{@{}lllll@{}}
\toprule
Country        & Target is NA & Target is 0 & Forecast is NA & Forecast is 0 \\ \midrule
Albania        & 49.47\%      & 0.00\%      & 33.59\%        & 0.00\%        \\
Austria        & 1.97\%       & 0.00\%      & 0.99\%         & 0.00\%        \\
Belgium        & 9.24\%       & 0.00\%      & 8.28\%         & 0.00\%        \\
Bulgaria       & 13.68\%      & 0.00\%      & 13.00\%        & 0.00\%        \\
Croatia        & 0.42\%       & 0.28\%      & 0.45\%         & 0.16\%        \\
Czech Republic & 0.54\%       & 0.00\%      & 0.36\%         & 0.00\%        \\
Denmark        & 2.36\%       & 0.00\%      & 1.82\%         & 0.00\%        \\
Estonia        & 0.91\%       & 0.00\%      & 0.50\%         & 0.00\%        \\
Finland        & 0.51\%       & 0.00\%      & 0.10\%         & 0.00\%        \\
France         & 0.90\%       & 0.00\%      & 0.25\%         & 0.00\%        \\
Germany        & 10.37\%      & 0.00\%      & 21.86\%        & 0.00\%        \\
Greece         & 0.95\%       & 0.00\%      & 0.28\%         & 0.00\%        \\
Hungary        & 1.39\%       & 0.00\%      & 1.08\%         & 0.00\%        \\
Ireland        & 12.81\%      & 0.25\%      & 37.17\%        & 0.00\%        \\
Italy          & 10.20\%      & 0.00\%      & 1.23\%         & 0.00\%        \\
Latvia         & 0.43\%       & 0.00\%      & 0.34\%         & 0.00\%        \\
Lithuania      & 0.79\%       & 0.03\%      & 0.75\%         & 0.00\%        \\
Luxembourg     & 5.88\%       & 0.00\%      & 2.63\%         & 0.01\%        \\
Macedonia      & 11.03\%      & 0.08\%      & 6.78\%         & 0.47\%        \\
Montenegro     & 1.74\%       & 1.02\%      & 0.57\%         & 1.01\%        \\
Netherlands    & 1.41\%       & 0.00\%      & 2.34\%         & 0.00\%        \\
Norway         & 3.32\%       & 0.00\%      & 2.36\%         & 0.00\%        \\
Poland         & 0.43\%       & 0.00\%      & 0.58\%         & 0.00\%        \\
Portugal       & 0.56\%       & 0.08\%      & 0.25\%         & 0.00\%        \\
Romania        & 2.79\%       & 0.00\%      & 0.82\%         & 0.00\%        \\
Serbia         & 3.36\%       & 0.01\%      & 3.71\%         & 0.00\%        \\
Slovakia       & 0.47\%       & 0.00\%      & 0.33\%         & 0.00\%        \\
Slovenia       & 0.93\%       & 0.03\%      & 0.86\%         & 0.00\%        \\
Spain          & 0.53\%       & 0.00\%      & 0.17\%         & 0.00\%        \\
Sweden         & 4.86\%       & 0.40\%      & 5.02\%         & 0.00\%        \\
Switzerland    & 8.34\%       & 0.00\%      & 0.44\%         & 0.00\%        \\
United Kingdom & 13.53\%      & 0.00\%      & 38.32\%        & 0.00\%        \\ \bottomrule
\end{tabular}
\end{table}

In general it is beneficial that the errors are taken care of in time, but this issue makes it harder to act upon the reported information on the portal. Also it would be beneficial to have, at least for the research community, a simple API for the data that is otherwise only accessible through screen scraping.

\section{Methodology}
The aim of this paper is not only to assess the utility of the new ENTSO-E platform, but to motivate publishing open access time series data in energy by building accurate models using the state-of-the-art in machine learning.

Point forecasts for time series are often provided by reliable statistical methods like ARMA and ARIMA. Our approach to forecasting is based on a relatively new machine learning method; Gradient Boosted Regression Trees (GBRTs) \cite{Friedman} were successful in a number of competitions in the past \cite{Gefcom2012i, Gergo}.

\subsection{Gradient Boosting Regression Trees}

Gradient boosting is an ensemble method responsible for combining weak
learners for higher model accuracy, as suggested by Friedman in 2000
\cite{friedman2001}. The predictor generated in gradient boosting is a linear
combination of weak learners, here tree models are utilized for this purpose.

%A sequence of models is built iteratively, and the final predictor is the
%weighted average of these predictors.
%Boosting generally results in an additive prediction function:
%
%\begin{equation} \label{eq:boost}
	%f^*(X) = \beta_0 + f_1(X_1)+ \ldots + f_p(X_p)
%\end{equation}
%
%In each turn of the iteration the ensemble calculates two set of weights:
%\begin{enumerate}
  %\item one for the current tree in the ensemble
  %\item one for each observation in the training dataset
%\end{enumerate}
%
%The rows in the training set are iteratively reweighted by upweighting
%previously misclassified observations.

A sequence of models is built iteratively, starting from a constant initial prediction $F_0$ (in a least squares regression task, the  mean of the target variable), and with each tree built on the residuals of the previous tree. To avoid overfitting, there is a shrinkage parameter (also called learning rate), which controls how much each tree contributes to the final model; this effectively slows learning down, yielding better performance at the expense of requiring more iterations.

The final prediction $F_n$ is calculated as follows:

\begin{equation} \label{eq:boost}
F_n(x) = F_0 + \sum_{i=1}^{n} \nu * h_i(x)
\end{equation}

Where $N$ is the number of individual trees built, $F_0$ is the initial prediction, $F_i$ is the prediction at stage $i$, $h$ is the prediction of an individual tree learner and $\nu \in (0,1]$ is the shrinkage parameter to reduce the effect of an individual tree in the sequence.

The general idea is to compute a sequence of simple trees, where each successive
tree is built for the prediction residuals of the preceding tree.
Each new base-learner is chosen to be maximally correlated with the negative
gradient of the loss function, associated with the whole ensemble.
This way the subsequent stages will focus harder on fitting these examples and
the resulting predictor is a linear combination of weak learners.

Utilizing boosting has many beneficial properties; various risk functions are
applicable, intrinsic variable selection is carried out, also resolves
multicollinearity issues, and works well with large number of features without
overfitting.

\subsection{Evaluation}

Model performance is measured using the Mean Absolute Percentage Error (MAPE) for point forecasts, a well-established error measure throughout the energy industry domain.
Probabilistic error rates were assessed using the Pinball loss function \cite{Koenker1978}, a well-known tool in both statistics and machine learning and quantile time series forecasting in particular \cite{Biau2011}. 

For a quantile forecast $q_a$ with $a/100$ as the target quantile, the pinball loss score $L$ is defined as:

\begin{equation} \label{eq:pinball}
    L(q_a, y) = \begin{cases}
		(1 - $a/100$) (q_a - y), &\text{if $y < q_a$};\\
		$a/100$ (y - q_a), &\text{if $y \ge q_a$}; 
		\end{cases}
\end{equation}

where $y$ is the observation used for verification, and a = 1, 2, \dots, 99.

The pinball loss function returns a value that can be interpreted as the accuracy of a quantile forecasting model. Being a loss function the lower rates are better.

\section{Framework structure}

\subsection{Module overview}
\begin{figure} \centering
\includegraphics[width=8cm]{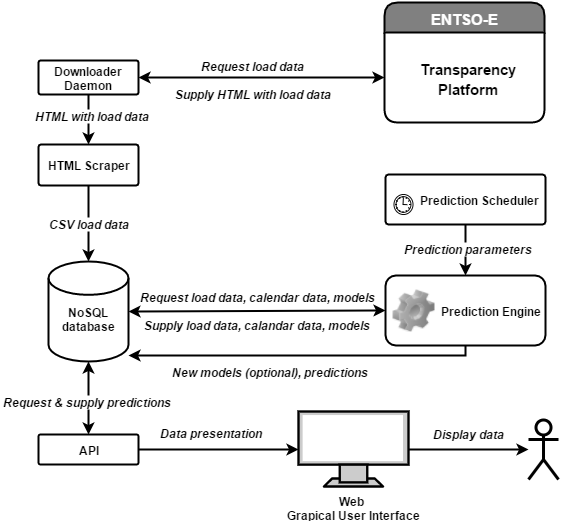}
\caption{Block diagram of framework modules}
\label{fig:blockdiag}
\end{figure}

The framework consists of modules responsible for scraping the data, storing it in a NoSQL database, building GBRT models and using the models to issue timely forecasts (see Figure \ref{fig:blockdiag}). The main output of the framework is electric load value predictions for the next $24$ hours for the selected countries. The prediction function launches $24$ subprocesses with prediction horizon set between $1$ and $24$ hours, it is also capable of parallel execution. Each subprocess consists of three main parts; data preparation, model building (if model rebuild is requested) and model application.

The end result of the prediction function is a dataset of $24$ predicted hourly load values ahead of the test time for the given country. This means that if the forecasting system is in continuous operation, there will be $24$ forecasts for each hour, with horizons decreasing from $24$ to $1$.

\subsection{Data preparation}
First, Vertical Load and total load values and the calendar of the target country are loaded from the NoSQL database. Total load means a load equal to generation and any imports deducting any exports and power used for energy storage. While Vertical Load equals the total amount of power flowing out of the transmission network to the distribution networks, to directly connected final customers or to the consuming part of generation\cite{emrwiki}. There is an issue that total load values are not available prior to $2015$, but in previous years substituting total load with Vertical Load data improves performance. This is a curious behaviour as Vertical Load and Total Load are inherently different concepts. Needless to say, when enough Total Load data has been acquired this tweak no longer will be necessary. For some countries, Total Load values are available in quarter-hourly or half-hourly granularity, so the data preparation module aggregates them to uniform hourly values. 

Next, feature vectors are created out of the load value observations. Calculating the feature vectors involves both calendar variables and lagged values. Do note, although desired at this point no weather data is available to be used.
Currently two lagged values are used; the load value one week before the predicted time slot and the last known load value (adjusted by the forecast horizon).
If a lagged load value happens to be missing in the training data that row is discarded entirely. It is crucial that no test row has missing values, so if a lagged load is missing the data preparation function imputes it using values from previous periods. If no lagged values are found for the past 6 weeks, the module switches to the more robust setting where no lagged load variables are used.

\subsection{Model building}
This part is executed only if the model rebuild is requested, during normal operations model rebuild is scheduled for every midnight.
There are two distinct models built: basic and advanced. Basic models are constructed on calendar-based features only, and are used as fall back method. Advanced models use lagged load features in addition meaning their performance is much better. The underlying algorithm used for both models is Gradient Boosted Regression Trees. For each country the basic version builds 50 trees with a depth of 5, while the up-scaled version has twice as many trees with larger depth. All generated models are saved in the NoSQL database along with their metadata.

\subsection{Model application}
In this step predictions for each test row are generated by applying the most recent model in the database that matches the country, horizon and model type (basic or advanced).

\section{Results}

\subsection{Model evaluation \& result comparison}
To justify the usage of a new approach in time series forecasting it is often benchmarked with an established methodology like ARMA on a set of time series in the respective domain \cite{Gergo}. In this particular application, if provided in the open access data, the country issued estimates of energy consumption are used to assess the relative accuracy of the GBRT methodology. These estimates are issued by the respective country as part of their transparency regulations. The framework provides 24 estimates for every given time point with a forecast horizon spanning between 1 and 24 hours. Day ahead forecasts are typical in energy industry so if otherwise not stated all evaluation was performed on the 24 hours horizon.

\begin{figure} \centering
\includegraphics[width=10cm]{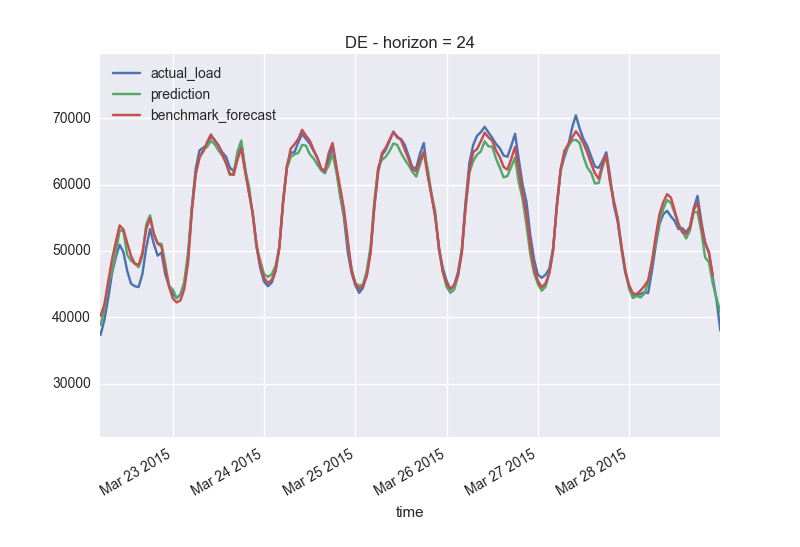}
\caption{Sample forecasted load, actual load \& benchmark for Germany in March 2015}
\label{fig:germany}
\end{figure}

Performance is measured using the Mean Absolute Percentage Error (MAPE) and model forecasts are compared to Actual Load values directly, and error measures are compared to country benchmark model forecasts as seen on the ENTSO-E Transparency Platform. Figure \ref{fig:germany} shows how different time series stack up in the sample period of late March.

% Please add the following required packages to your document preamble:
% \usepackage{booktabs}
\begin{table}[]
\centering
\caption{Measured error by country \& month}
\label{tab:mapecm}
\begin{tabular}{@{}lrrrrrrrr@{}}
\toprule
\multicolumn{1}{c}{} & \multicolumn{1}{c}{Mar} & \multicolumn{1}{c}{Apr} & \multicolumn{1}{c}{May} & \multicolumn{1}{c}{Jun} & \multicolumn{1}{c}{Jul} & \multicolumn{1}{c}{Aug} & \multicolumn{1}{c}{Sep} & \multicolumn{1}{c}{Oct} \\ \midrule
BE                   & 4.481                   & 3.474                   & 5.232                   & 3.347                   & 5.309                   & 4.817                   & 3.846                   & 4.335                   \\
CZ                   & 3.409                   & 2.958                   & 3.395                   & 2.275                   & 4.587                   & 4.242                   & 3.582                   & 5.564                   \\
DE                   & 5.478                   & 7.171                   & 6.222                   & 3.146                   & 5.606                   & 4.288                   & 3.822                   & 5.981                   \\
FI                   & 4.232                   & 2.738                   & 3.347                   & 3.803                   & 3.339                   & 2.624                   & 5.437                   & 5.841                   \\
FR                   & 6.237                   & 3.483                   & 4.664                   & 3.384                   & 5.485                   & 4.954                   & 4.461                   & 8.671                   \\
HU                   & 2.603                   & 3.278                   & 3.270                   & 3.102                   & 6.107                   & 9.859                   & 4.577                   & 4.053                   \\
IT                   & 4.279                   & 8.589                   & 5.891                   & 4.961                   & 6.191                   &                         &                         &                         \\
NO                   & 5.014                   & 3.618                   & 9.551                   & 4.056                   & 5.262                   & 4.907                   & 7.941                   & 5.554                   \\
PL                   & 3.181                   & 3.597                   & 4.505                   & 2.314                   & 3.953                   & 3.929                   & 2.985                   & 3.814                   \\
SK                   & 3.567                   & 3.610                   & 3.030                   & 2.281                   & 5.069                   & 3.711                   & 3.407                   & 3.570                   \\ \bottomrule
\end{tabular}
\end{table}

Table \ref{tab:mapecm} summarizes measured MAPE over the test period for each country. The GBRT framework generally performs well, but accuracy varies from country to country, the best being Czech Republic, Poland, Slovakia and to some extent Hungary. The model coped very well with the different seasons; all 3 have very similar mean error rates. Do note, that there was no Actual Load data published for Italy from August and onwards. 

Performance measurements are put into context when looking at the country provided estimates used as benchmark here (see Table \ref{tab:bmcm}). Observed MAPE rates are typically lower here, with the sole exception of Czech Republic. Cases where our GBRT framework offers comparable results include Czech Republic, Hungary and to a lesser extent Germany and Slovakia. Scandinavian and Benelux countries operate with extremely low error rates. At this point, very little is known of the models and data used by any of the countries. Though country estimates could potentially (and should) incorporate additional information besides historical and calendar-related data such as weather forecasts, longer spanning historical data, customer data, market-related data. Again, providing at least partial open-access to this data could greatly boost development in this field.  

% Please add the following required packages to your document preamble:
% \usepackage{booktabs}
\begin{table}[]
\centering
\caption{Benchmark error by country \& month}
\label{tab:bmcm}
\begin{tabular}{@{}lrrrrrrrr@{}}
\toprule
\multicolumn{1}{c}{} & \multicolumn{1}{c}{Mar} & \multicolumn{1}{c}{Apr} & \multicolumn{1}{c}{May} & \multicolumn{1}{c}{Jun} & \multicolumn{1}{c}{Jul} & \multicolumn{1}{c}{Aug} & \multicolumn{1}{c}{Sep} & \multicolumn{1}{c}{Oct} \\ \midrule
BE                          & 1.848                   & 1.964                   & 1.694                   & 1.377                   & 2.139                   & 3.426                   & 1.434                   & 1.709                   \\
CZ                          & 5.429                   & 5.898                   & 4.638                   & 5.360                   & 6.532                   & 5.935                   & 6.897                   & 5.374                   \\
DE                          & 2.886                   & 3.744                   & 3.299                   & 5.958                   & 4.711                   & 4.042                   & 2.962                   & 3.617                   \\
FI                          & 1.368                   & 1.761                   & 1.657                   & 1.840                   & 1.287                   & 1.405                   & 2.934                   & 1.632                   \\
FR                          & 1.984                   & 1.690                   & 1.640                   & 1.494                   & 1.644                   & 1.712                   & 1.540                   & 1.701                   \\
HU                          & 3.159                   & 4.371                   & 4.294                   & 3.698                   & 4.577                   & 4.864                   & 4.551                   & 4.468                   \\
IT                          & 1.617                   & 1.980                   & 1.633                   & 1.683                   & 2.043                   &                         &                         &                         \\
NO                          & 0.092                   & 0.357                   & 0.165                   & 0.092                   & 0.506                   & 2.102                   & 4.542                   & 1.990                   \\
PL                          & 1.643                   & 1.915                   & 1.585                   & 1.560                   & 2.176                   & 2.338                   & 2.147                   & 1.725                   \\
SK                          & 2.152                   & 2.318                   & 2.267                   & 2.199                   & 2.773                   & 2.258                   & 2.560                   & 2.423                   \\ \bottomrule
\end{tabular}
\end{table}

\begin{figure} \centering
\includegraphics[width=8cm]{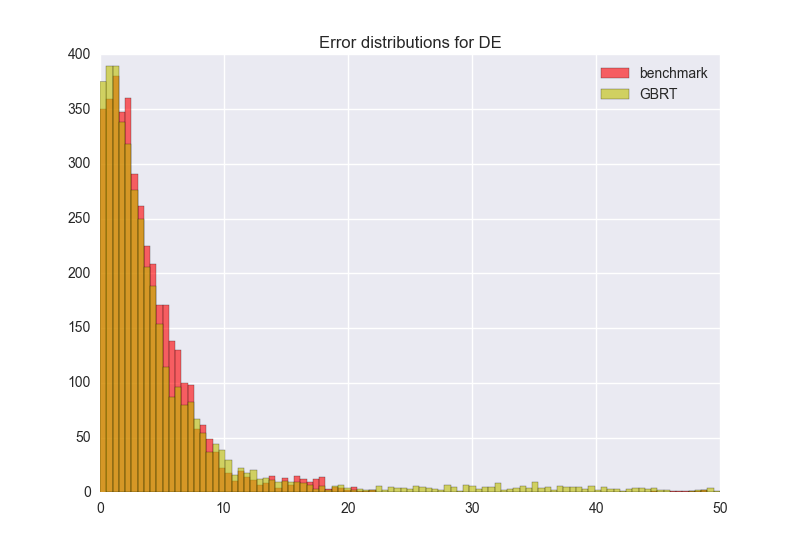}
\caption{Model output and benchmark error distributions for Germany}
\label{fig:errdist}
\end{figure}

Looking at the error statistics in Table \ref{tab:errdist} one can see that all error distributions are naturally right skewed suggesting a distribution close to log-normal, meaning the GBRT powered framework more frequently commits smaller errors than higher ones. The framework performance is rather robust over countries, major error percentiles are well aligned with the distribution tail having varying length. Figure \ref{fig:errdist} shows sample distributions for Germany comparing framework and benchmark errors. The GBRT errors demonstrate a much longer tail indicating that some patterns or information might be missing from this representation. On the other hand errors close to zero are also more frequent in the model output.

% Please add the following required packages to your document preamble:
% \usepackage{booktabs}
\begin{table}[]
\centering
\caption{MAPE error statistics by country}
\label{tab:errdist}
\begin{tabular}{@{}lrrrrrrr@{}}
\toprule
\multicolumn{1}{c}{} & \multicolumn{1}{c}{mean} & \multicolumn{1}{c}{std} & \multicolumn{1}{c}{min} & \multicolumn{1}{c}{25\%} & \multicolumn{1}{c}{50\%} & \multicolumn{1}{c}{75\%} & \multicolumn{1}{c}{max} \\ \midrule
BE                   & 4.325                    & 5.265                   & 0.002                   & 1.247                    & 2.714                    & 5.066                    & 47.422                  \\
CZ                   & 3.904                    & 6.116                   & 0                       & 0.832                    & 1.97                     & 4.144                    & 53.674                  \\
DE                   & 5.17                     & 8.015                   & 0.001                   & 1.267                    & 2.747                    & 5.307                    & 116.77                  \\
FI                   & 3.98                     & 5.612                   & 0                       & 0.994                    & 2.103                    & 4.211                    & 42.351                  \\
FR                   & 5.321                    & 8.482                   & 0                       & 1.138                    & 2.575                    & 5.404                    & 76.028                  \\
HU                   & 4.924                    & 6.318                   & 0.001                   & 0.994                    & 2.476                    & 5.944                    & 39.584                  \\
IT                   & 5.924                    & 10.815                  & 0.001                   & 1.154                    & 2.588                    & 5.724                    & 88.475                  \\
NO                   & 5.567                    & 9.122                   & 0                       & 1.185                    & 2.604                    & 5.315                    & 79.224                  \\
PL                   & 3.571                    & 6.15                    & 0.002                   & 0.754                    & 1.724                    & 3.793                    & 59.342                  \\
SK                   & 3.607                    & 4.619                   & 0.001                   & 0.911                    & 2.061                    & 4.326                    & 40.367                  \\ \bottomrule
\end{tabular}
\end{table}

The framework provides 24 estimates for every given time point, shorter horizons often produce better results as underlined by Table \ref{tab:horizonstats}. Forecast errors between 1 and 24 hours length typically draw an arc, with higher error rates in the middle and the lowest on the edges. But the distribution is far from symmetrical; a 1-length horizon typically yields approximately 12\% lower error rates than its 24 hour counterpart.

\begin{table}[]
\centering
\caption{Measured MAPE by horizon \& country}
\label{tab:horizonstats}
\begin{tabular}{@{}lllllllllll@{}}
\toprule
\multicolumn{1}{c}{} & \multicolumn{1}{c}{BE}   & \multicolumn{1}{c}{CZ}   & \multicolumn{1}{c}{DE}   & \multicolumn{1}{c}{FI}   & \multicolumn{1}{c}{FR}   & \multicolumn{1}{c}{HU}   & \multicolumn{1}{c}{IT}   & \multicolumn{1}{c}{NO}   & \multicolumn{1}{c}{PL}   & \multicolumn{1}{c}{SK}   \\ \midrule
1                    & \multicolumn{1}{r}{3.70} & \multicolumn{1}{r}{3.71} & \multicolumn{1}{r}{4.47} & \multicolumn{1}{r}{3.44} & \multicolumn{1}{r}{4.79} & \multicolumn{1}{r}{4.82} & \multicolumn{1}{r}{5.27} & \multicolumn{1}{r}{4.34} & \multicolumn{1}{r}{3.43} & \multicolumn{1}{r}{3.43} \\
2                    & \multicolumn{1}{r}{3.96} & \multicolumn{1}{r}{3.79} & \multicolumn{1}{r}{4.73} & \multicolumn{1}{r}{3.60} & \multicolumn{1}{r}{4.86} & \multicolumn{1}{r}{4.88} & \multicolumn{1}{r}{5.92} & \multicolumn{1}{r}{4.54} & \multicolumn{1}{r}{3.49} & \multicolumn{1}{r}{3.47} \\
3                    & \multicolumn{1}{r}{4.23} & \multicolumn{1}{r}{3.83} & \multicolumn{1}{r}{4.86} & \multicolumn{1}{r}{3.72} & \multicolumn{1}{r}{4.91} & \multicolumn{1}{r}{4.85} & \multicolumn{1}{r}{6.15} & \multicolumn{1}{r}{4.73} & \multicolumn{1}{r}{3.53} & \multicolumn{1}{r}{3.53} \\
4                    & \multicolumn{1}{r}{4.37} & \multicolumn{1}{r}{3.91} & \multicolumn{1}{r}{4.99} & \multicolumn{1}{r}{3.78} & \multicolumn{1}{r}{4.97} & \multicolumn{1}{r}{4.81} & \multicolumn{1}{r}{6.36} & \multicolumn{1}{r}{4.88} & \multicolumn{1}{r}{3.57} & \multicolumn{1}{r}{3.55} \\
5                    & \multicolumn{1}{r}{4.47} & \multicolumn{1}{r}{3.93} & \multicolumn{1}{r}{5.11} & \multicolumn{1}{r}{3.86} & \multicolumn{1}{r}{5.03} & \multicolumn{1}{r}{4.79} & \multicolumn{1}{r}{6.53} & \multicolumn{1}{r}{4.95} & \multicolumn{1}{r}{3.62} & \multicolumn{1}{r}{3.59} \\
6                    & \multicolumn{1}{r}{4.52} & \multicolumn{1}{r}{3.96} & \multicolumn{1}{r}{5.21} & \multicolumn{1}{r}{3.91} & \multicolumn{1}{r}{5.09} & \multicolumn{1}{r}{4.81} & \multicolumn{1}{r}{6.69} & \multicolumn{1}{r}{5.08} & \multicolumn{1}{r}{3.65} & \multicolumn{1}{r}{3.60} \\
7                    & \multicolumn{1}{r}{4.68} & \multicolumn{1}{r}{3.98} & \multicolumn{1}{r}{5.31} & \multicolumn{1}{r}{3.97} & \multicolumn{1}{r}{5.11} & \multicolumn{1}{r}{4.81} & \multicolumn{1}{r}{6.77} & \multicolumn{1}{r}{5.10} & \multicolumn{1}{r}{3.66} & \multicolumn{1}{r}{3.65} \\
8                    & \multicolumn{1}{r}{4.75} & \multicolumn{1}{r}{3.96} & \multicolumn{1}{r}{5.37} & \multicolumn{1}{r}{4.02} & \multicolumn{1}{r}{5.17} & \multicolumn{1}{r}{4.82} & \multicolumn{1}{r}{6.82} & \multicolumn{1}{r}{5.16} & \multicolumn{1}{r}{3.66} & \multicolumn{1}{r}{3.61} \\
9                    & \multicolumn{1}{r}{4.78} & \multicolumn{1}{r}{4.02} & \multicolumn{1}{r}{5.41} & \multicolumn{1}{r}{4.04} & \multicolumn{1}{r}{5.19} & \multicolumn{1}{r}{4.82} & \multicolumn{1}{r}{6.76} & \multicolumn{1}{r}{5.24} & \multicolumn{1}{r}{3.63} & \multicolumn{1}{r}{3.62} \\
10                   & \multicolumn{1}{r}{4.86} & \multicolumn{1}{r}{4.01} & \multicolumn{1}{r}{5.44} & \multicolumn{1}{r}{4.05} & \multicolumn{1}{r}{5.21} & \multicolumn{1}{r}{4.83} & \multicolumn{1}{r}{6.71} & \multicolumn{1}{r}{5.28} & \multicolumn{1}{r}{3.62} & \multicolumn{1}{r}{3.62} \\
11                   & \multicolumn{1}{r}{4.87} & \multicolumn{1}{r}{4.02} & \multicolumn{1}{r}{5.45} & \multicolumn{1}{r}{4.09} & \multicolumn{1}{r}{5.22} & \multicolumn{1}{r}{4.81} & \multicolumn{1}{r}{6.70} & \multicolumn{1}{r}{5.37} & \multicolumn{1}{r}{3.64} & \multicolumn{1}{r}{3.60} \\
12                   & 4.95                     & 4.04                     & 5.46                     & 4.10                     & 5.24                     & 4.77                     & 6.89                     & 5.42                     & 3.63                     & 3.62                     \\
13                   & 4.92                     & 4.06                     & 5.51                     & 4.09                     & 5.23                     & 4.78                     & 6.88                     & 5.44                     & 3.61                     & 3.62                     \\
14                   & 5.07                     & 4.02                     & 5.46                     & 4.15                     & 5.25                     & 4.75                     & 6.96                     & 5.43                     & 3.62                     & 3.63                     \\
15                   & 5.01                     & 4.03                     & 5.48                     & 4.14                     & 5.24                     & 4.74                     & 6.93                     & 5.47                     & 3.65                     & 3.64                     \\
16                   & 4.98                     & 4.04                     & 5.45                     & 4.14                     & 5.27                     & 4.76                     & 7.02                     & 5.55                     & 3.61                     & 3.60                     \\
17                   & 4.90                     & 4.03                     & 5.43                     & 4.17                     & 5.29                     & 4.77                     & 6.98                     & 5.51                     & 3.62                     & 3.61                     \\
18                   & 4.88                     & 3.99                     & 5.39                     & 4.16                     & 5.28                     & 4.76                     & 6.97                     & 5.51                     & 3.64                     & 3.61                     \\
19                   & 4.78                     & 4.00                     & 5.38                     & 4.10                     & 5.33                     & 4.78                     & 6.89                     & 5.51                     & 3.67                     & 3.63                     \\
20                   & 4.67                     & 3.96                     & 5.32                     & 4.11                     & 5.38                     & 4.80                     & 6.85                     & 5.51                     & 3.65                     & 3.61                     \\
21                   & 4.65                     & 3.99                     & 5.29                     & 4.06                     & 5.33                     & 4.83                     & 6.73                     & 5.46                     & 3.67                     & 3.58                     \\
22                   & 4.54                     & 3.99                     & 5.28                     & 4.04                     & 5.33                     & 4.87                     & 6.60                     & 5.49                     & 3.64                     & 3.58                     \\
23                   & 4.46                     & 3.93                     & 5.23                     & 4.03                     & 5.29                     & 4.89                     & 6.31                     & 5.49                     & 3.58                     & 3.60                     \\
24                   & 4.33                     & 3.90                     & 5.17                     & 3.98                     & 5.32                     & 4.92                     & 5.92                     & 5.57                     & 3.57                     & 3.61                     \\ \bottomrule
\end{tabular}
\end{table}

\begin{figure} \centering
\includegraphics[width=9cm]{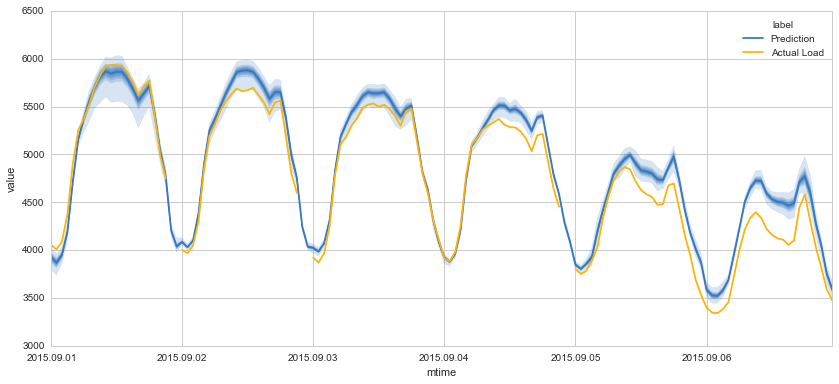}
\caption{Probabilistic forecasts for Hungary in the first week of September}
\label{fig:probfore}
\end{figure}

Furthermore we investigate the forecasting of a probability distribution of the target variable in contrast to forecasting a single value. Probabilistic forecasts provide more detailed insight for stakeholders offering immediate business value. For this purpose we utilize the quantile regression capabilities of the GBRT model and estimate target variable distribution over the major deciles. Probabilistic forecasts were experimented with Hungary only for the month of September, training the model on the period of January 2012 - August 2015. Figure \ref{fig:probfore} shows a sample of the estimated time series distribution along with the Actual Load. The quantile intervals here are relatively narrow on most days, with slight bias issues in some cases. 

\begin{figure} \centering
\includegraphics[width=7cm]{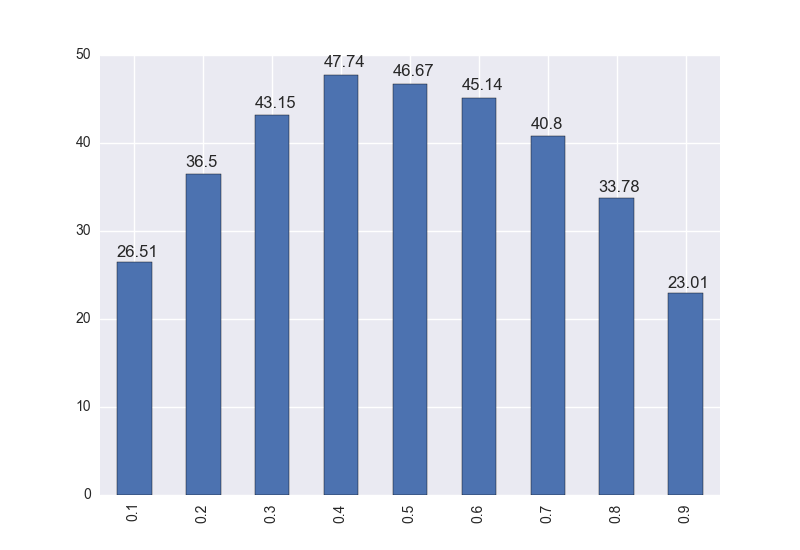}
\caption{Distribution of pinball loss over the quantiles}
\label{fig:pinballdist}
\end{figure}

Figure \ref{fig:pinballdist} shows the probabilistic Pinball Loss distribution over the quantiles. The distribution is almost symmetrical with higher rates observed close to the median. The average pinball loss measured for this setting is $38.144$. Unfortunately with only point forecasts published on the portal there is no direct way to compare the loss to country estimates. We look forward to probabilistic forecasts being released in the near future as they provide more detailed insight thus generating business value.  

\begin{figure} \centering
\includegraphics[width=8cm]{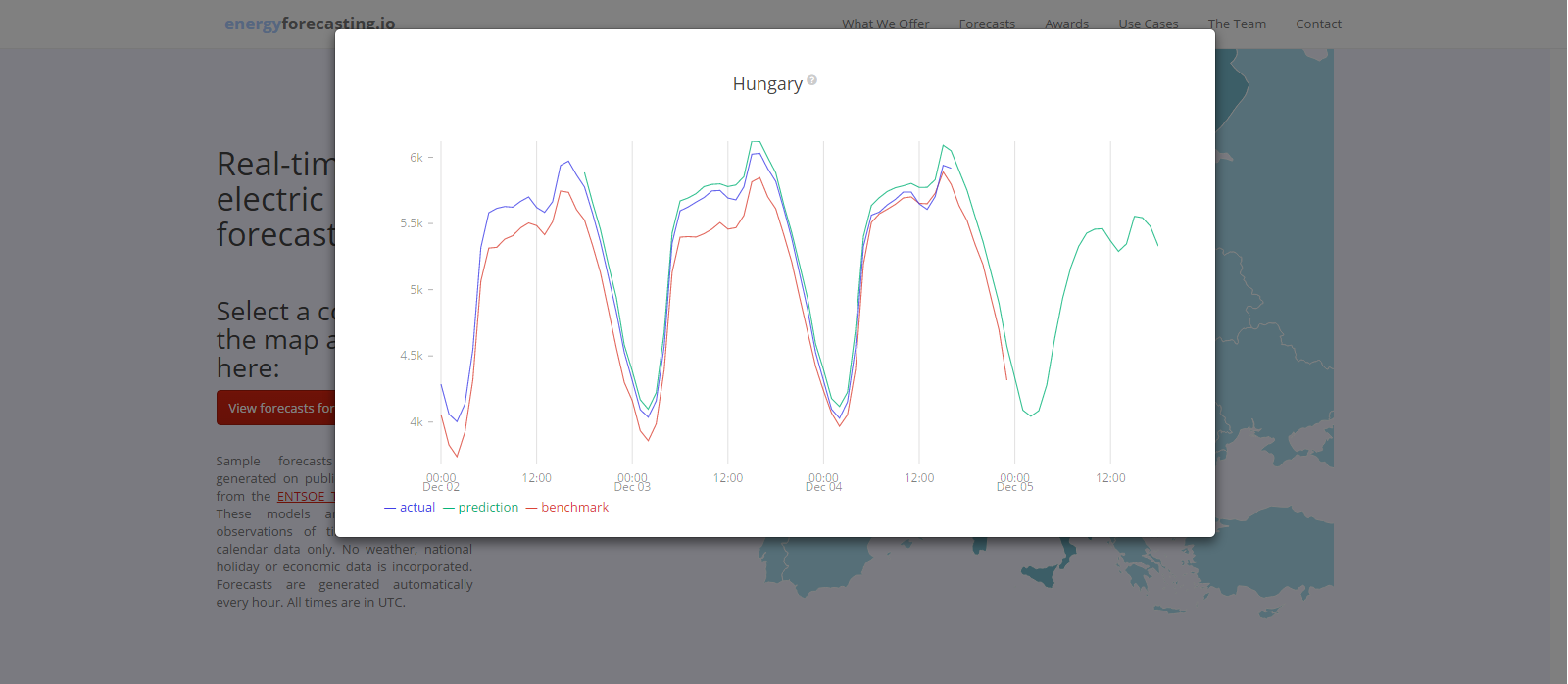}
\caption{Displaying forecasts on the web directly from the API}
\label{fig:efiohun}
\end{figure}

\section{Future work \& Conclusions}

The framework presented in this paper is functional and fully operational, yet there are still plenty of ideas to further extend it. As mentioned in the paper, at the moment no external data sources are used. Previous experience shows that incorporating country specific weather forecasts can greatly boost prediction accuracy and provide richer data sources encapsulated in the framework. ENTSOG integration, the gas counterpart of ENTSO-E, is also a reasonable next step as gas usage is even more weather dependent than electricity. To further justify the usage of GBRT in time series forecasting there is a need to benchmark it to an established methodology like ARMA on a relatively large set of time series in the energy domain in a rolling fashion. Country specific load data is particularly suitable for this purpose.

Benchmarking is not the sole purpose of this paper, as it deals with the data analysis process as a whole from ETL to data preparation (ENTSO-E integration, data cleansing and data auditing), through modeling and result presentation (see Figure \ref{fig:efiohun}) with an API providing parametric forecasts for a number of countries in Europe. It will hopefully contribute to the agenda of releasing more open access data to the public that can be put to good use in the future.

%\section{Future work}
%\begin{enumerate}
	%\item In this paper we propose a framework for automated forecasting of energy-related time series using open access data from European Network of Transmission System Operators for Electricity and \textbf{Gas} (ENTSO-E and \textbf{ENTSOG} respectively)
	%\item The framework provides forecasts for various European countries using a 3rd party API for weather data
	%\item incorporate it into our \textbf{web based API} that provides quick and easy access to reliable forecasts
  %\item It is also very interesting to see if available weather data results in more reliable forecasts. 
  %\item To justify the usage of GBRT in time series forecasting we often benchmark it to an established methodology like \textbf{ARMA} on a relatively large set of time series in the energy domain in a rolling fashion.
%\end{enumerate}

% can use a bibliography generated by BibTeX as a .bbl file
% BibTeX documentation can be easily obtained at:
% http://www.ctan.org/tex-archive/biblio/bibtex/contrib/doc/
% The IEEEtran BibTeX style support page is at:
% http://www.michaelshell.org/tex/ieeetran/bibtex/
%\bibliographystyle{IEEEtran}
% argument is your BibTeX string definitions and bibliography database(s)
%\bibliography{IEEEabrv,../bib/paper}
%
% <OR> manually copy in the resultant .bbl file
% set second argument of \begin to the number of references
% (used to reserve space for the reference number labels box)

% that's all folks
\end{document}